\documentclass[twocolumn,showpacs,preprintnumbers,amsmath,amssymb,aps,prl,nobibnotes]{revtex4}

\usepackage{graphicx}
\usepackage{dcolumn}
\usepackage{bm}

\begin{document}

\preprint{APS/123-QED}

\title{Nonlocal effects on the resonant dielectric surface\\ observed by near-field optical probing in mid-infrared range}

\author{Dmitry~Kazantsev}
\email{kaza@itep.ru}
\altaffiliation{The experiment was carried out
in Max-Planck institute of Biochemistry, Martinsried, Germany}

\affiliation{
Institute of Theoretical and Experimental Physics,
B.Cheremushkinskaya st., 25, Moscow, 117218, Russia.
}%

\date{\today}

\begin{abstract}
The local electromagnetic field distribution over the dielectric
(SiC) surface illuminated by mid-IR light of frequency close to the
lattice resonance was directly mapped with an apertureless
scattering near-field optical microscope. Half of the sample surface
was covered with a metal layer, in which some small $(0.2-10~\mu m)$
holes were formed. It was found, that due to the eliminating of
collective surface polarization effects, the amplitude of the
electromagnetic field becomes several times higher over such holes
than over infinitely open surface of the same dielectric.
\end{abstract}

\maketitle

Surface states, containing the mixture of strongly interacting
electromagnetic field and lattice oscillations at the solid surface
(surface phonon-polariton states, SPP~\cite{SPP_textbook}) attract
the interest as possible base of optoelectronic microdevices
~\cite{OptoMicroDevProposed_Hillenbr},
~\cite{OptoMicroDevProposed_Keilmann}, as well as surface plasmon
polariton phenomena~\cite{PlasmMicroDevProposed_Ebbessen}. Surface
polariton effects can be also utilized for the spectroscopy of
molecules adsorbed on the substrate~\cite{Anderson_APL2003}, if any
field enhancement on subwavelength-size surface features takes
place, like Raman scattering enhancement~\cite{SERS1},~\cite{SERS2}
on rough metal surface due to surface plasmon excitations.

Collective electromagnetic excitations in polar crystals, caused by
the resonant displacement of charged atoms in the presence of
electromagnetic fields, have been investigated since 1908
~\cite{Mie_NaCl}. Maxwell equations for the radiation must be solved
together with the equations of motion for the charged atoms in the
lattice. This task has been solved for bulk isotropic media
\cite{LST_1941},\cite{Born_book_1954},\cite{Barron_1961} and for
semi-infinite solid (it is well presented
in~\cite{Rup_Engm_1970},\cite{Mills_Burst_1974}). The components of
the E-field are presented as plane wave term $E_{j}=E_{0j}\cdot
e^{-\sigma_{z}(\omega)z}\cdot e^{ik_{x}(\omega)x}\cdot
e^{ik_{y}(\omega)y}$  with $j\in\{x,y,z\}$. For the light frequency
at the reststrahlen band (near lattice resonance, between
frequencies of purely transversal $\omega_{T}$ and purely
longitudinal $\omega_{L}$ modes in bulk) the solution is known as
the SPP mode. For cubic crystals the lateral SPP wavevector can be
expressed analytically as
\begin{equation}
k_{x,y}(\omega)=\frac{\omega}{c}\sqrt{\frac{\varepsilon_{v}\varepsilon_{lat}(\omega)}
{\varepsilon_{v}+\varepsilon_{lat}(\omega)}},
 \varepsilon_{v}\equiv1,\label{eq:k_xy}\\
\end{equation}\\
and the field decay in normal direction is described by\\
\begin{equation}
\sigma_{z(beneath)}(\omega)=i\frac{\omega}{c}
\frac{\varepsilon_{lat}(\omega)}{\sqrt{\varepsilon_{v}+\varepsilon_{lat}(\omega)}}
\label{eq:sigma_z}
\end{equation}
where  $\varepsilon_{lat}(\omega)$ is a frequency-dependent
dielectric function, which can be expressed from experimentally
observed bulk phonon polariton frequencies $\omega_{L}$ and
$\omega_{T}$
\begin{equation}
\varepsilon_{lat}(\omega)=\varepsilon_{lat}(\infty)
(1+\frac{\omega^{2}_{L}+\omega^{2}_{T}}{\omega^{2}_{T}-\omega^{2}-i\omega\Gamma})
\label{eq:eps_omega}
\end{equation}
The consideration presented above assumes the collective nature of
the SPP oscillations on the infinite surface. The wave vector
conservation law suppresses effective excitation of the SPP by the
external radiation in this ideal case. This symmetry rule can be
lifted by any discontinuities of the surface electromagnetic
properties, in particular by the sharp edge(s) of the metal
mask~\cite{Sievers_APL32_edge_launch} formed on the surface. It was
hardly possible to prove the validity of description of the SPP
excitation near finite-size mask structures by the direct probing of
the field distribution in sub-wavelength scale, until
SNOM~\cite{SNOM_first} was invented. The point-like tip of SNOM has
no translational symmetry, and therefore no symmetry restrictions.
The scattering type (s-SNOM~\cite{s_SNOM_first}) setup is a kind of
SNOM in which the light scattered by the tapping AFM
~\cite{firstAFM} tip is collected. The s-SNOM allows amplitude and
phase~\cite{Hillenbr_Phase_PRL_2000} mapping of the electromagnetic
phenomena in close vicinity of the surface with a spatial resolution
at some cases~\cite{SNOM_1nm} as sharp as 1 nm, regardless to the
wavelength. The tip scatters some fraction of the SPP field to
external space, and the amplitude of scattered wave can be in
dipolar approximation described~\cite{Keilmann_Alpha_eff_2004} by
\begin{equation}
E_{sc}\propto \alpha_{t}(\varepsilon_{s}(x,y), z_{ts})\cdot E_{loc}
\label{eq:Esc_alpha}
\end{equation}
where $E_{loc}$ is the field at the tip location. The effective
polarizability $\alpha_{t}(\varepsilon_{s}(x,y), z_{ts})$ of the tip
is a nonlinear function of the tip-sample separation  and strongly
depends on the local dielectric value of the sample.

In our experiment we investigated the configuration of the surface
electromagnetic field excited on the surface of resonant media by
the external radiation. We used the SiC crystal sample with c-axis
normal to the surface. The laser frequency range (880-936~cm$^{-1}$)
was close to the lattice resonance of SiC ($\nu_{T}$=796~cm$^{-1}$,
$\nu_{L}$=968~cm$^{-1}$), so that one could expect an effective SPP
excitation caused by the laser beam. The sample surface was covered
with 120 nm Au film, which is thick enough to be completely opaque
for the light wave used. To define small ($0.2-10~\mu$m) holes in
the metal layer, a salt dust mask was used in the evaporation
process, so that some "lakes" of SiC remained open in the Au mask
after the salt was washed away. Half of the sample surface was left
uncover, so that we are able to compare the light scattering
response on gold, on clean SiC surface far from the metal edge, and
on SiC surface at small holes in Au layer.

\begin{figure}
\includegraphics[width=\columnwidth]{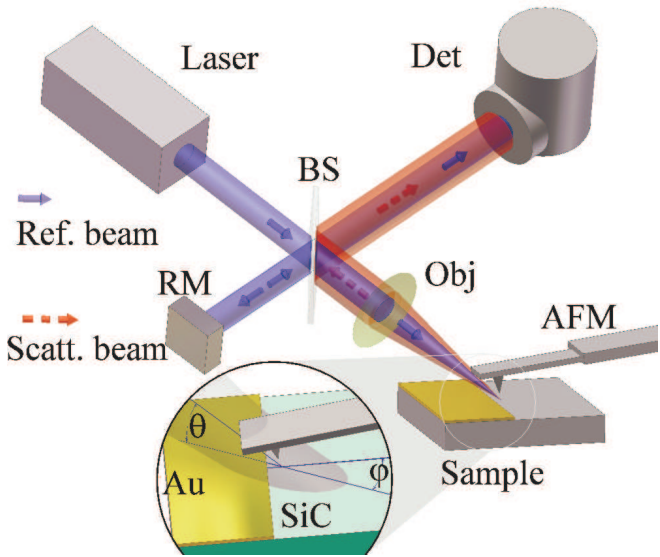}
\caption{\label{fig:setup} Experimental setup geometry. Laser -
$CO_2$ tunable laser ($880-936~cm^{-1}$), BS - interferometer
beamsplitter, RM - reference beam mirror, Obj - objective to focus
laser beam onto sample and tip and to collect scattered light, Det -
mid-IR photodetector.}
\end{figure}

The experimental setup, based on s-SNOM instrument~\cite{MPI_SNOM},
is shown in Fig.~\ref{fig:setup}. The beam of the tunable
$^{13}C^{16}O_2$ laser is focused onto the sample, so that the
Pt-coated Si tip of tapping-mode AFM is in the center of focal spot.
The shape of the focal spot could be described by soft (Gaussian)
function of some hundreds micrometers width, and the phase fronts of
the illumination light were proven to be flat across whole spot.
That means, that the sample illumination could be well described by
the plane wave with field amplitude $E_{las0}$. Thus, the direction
of the exciting light is well defined in our experiment by azimuthal
angle $\varphi$ and incidence angle $\theta=30^{\circ}$ to the
sample surface. The light scattered by the tip is collected back to
the Michelson interferometer, and the variations of the detector
signal $I_{det}(t)$ are demodulated at higher harmonic components of
the tip tapping frequency $\Omega$, utilizing the nonlinearity of
$\alpha_{t}( \varepsilon(x,y),z_{ts})$ dependency on $z_{ts}$. The
averaging of demodulated signal $I^{(n\Omega)}_{det}$ over reference
beam phase~\cite{MPI_Interfr_Detect} gives us full knowledge
concerning amplitude and phase of scattered radiation variations.

\begin{figure}
\includegraphics[width=\columnwidth]{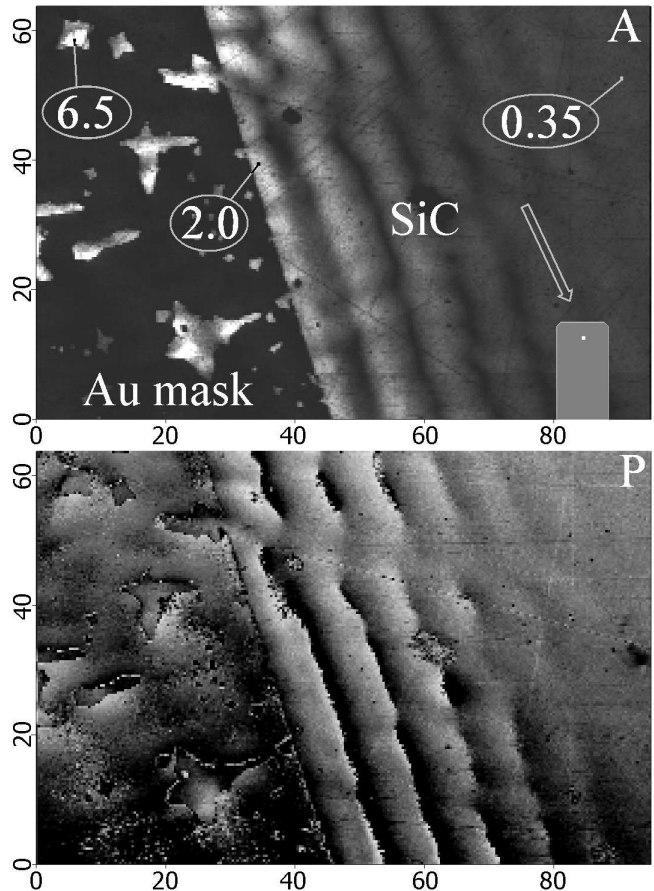}
\caption{\label{fig:expMap} Tip scattering amplitude (A) and phase
(P) images of the SiC surface recorded at frequency 936~cm$^{-1}$.
The edge of the Au mask, with small holes in it, can be seen. The
"pump" light direction (shown by arrow) is parallel to the metal
film edge. The cantilever beam (schematically shown at the image) is
mainly parallel to the light direction. The values in the ovals
represent light scattering amplitude at the marked points. The scan
area dimensions are $95\times65~\mu m$}
\end{figure}

The near field scattering component amplitude/phase map is shown in
Fig.~\ref{fig:expMap}. The spatial resolution is as sharp as at
least 200 nm (judging by 10\%-90\% step) in topography, and also
180-200 nm in the optical response. That shows not only high lateral
resolution of the instrument used, but also the sharpness of the
mask features on the sample.

We have shown in our previous work~\cite{APL_2005} that, in our
particular case (media, tip, tapping amplitude and light frequency),
the signal $I^{(n\Omega)}_{det}$ detected in the experiment is in
fact proportional to the E-field above the surface and represents
therefore the local polarization of the surface being mapped:

\begin{equation}
I^{(2\Omega)}_{det}
\propto\alpha^{(2\Omega)}_{eff}(\varepsilon_s(x,y))\cdot
E_{loc}(x,y) \label{eq:Idet_2omega}
\end{equation}




Here complex factor $\alpha^{(2\Omega)}_{eff}( \varepsilon_s(x,y))$
replaced $\alpha_{t}( \varepsilon_s(x,y),z_{ts})$, is a local
property of surface point, determined by the tip motion, scattering
z-dependency nonlinearity and the photocurrent demodulation at the
2nd harmonic component of the tip tapping frequency $\Omega$.

We have shown also that the local field $E_{loc}$, in turn, could be
expressed as sum of incident radiation field $E_{las}$ and some term
proportional to the amplitude $E^{\prime}_{spp}$ of [running] SPP
wave(s), which was launched somewhere far from the tip location:

\begin{equation}
E_{loc}(x,y)= (E_{las}+E^{\prime}_{spp}) = (E_{las0}+\eta
P_{0spp})e^{i \omega t}\label{eq:E_sum}
\end{equation}

where $\eta$ is a constant to convert polarization $P_{spp}$ below
the surface to the field $E^{\prime}_{spp}$ above the surface by
continuity conditions. Strictly speaking, Eqs.
(\ref{eq:k_xy}-~\ref{eq:eps_omega}) for anisotropic SiC are a bit
more complicated than for cubic crystals (and we have taken that
into account), but the difference in $\varepsilon _a$, $\varepsilon
_c$ is just a few
percent~\cite{Eps_SiC_Landolt_B},~\cite{Eps_SiC_PRB_1997} for SiC.
Therefore, even if one neglects the anisotropy completely, it
results in no more than 1-2\% error in our case.

As one can see on the map (Fig.\ref{fig:expMap}), the E-field
amplitude above SiC surface within small openings in the metal mask
is several times (we observed up to 20 times for some frequencies)
higher than amplitude over open surface of the SiC far from the
metal edge. That means the amplitude of local E-field at any surface
point does not depend only on local amplitude/phase of incident
light and on local sample dielectric constant at given light
frequency. It indicates clearly, that also the presence of the
resonant media far beyond the tip-sample near-field interaction
region (of dimensions known to be about tip radius) plays
significant role, so that excluding most of the surface around (with
the opaque metal mask) leads to the dramatic increase of the local
response at the point being investigated. In other words, all
surface points of SiC lattice, driven by the external
electromagnetic radiation, deliver their response to the point of
observation as SPP wave. For the infinite open surface, the role of
that polarization of neighbors is (in average) destructive in
formation of the local field.

We can demonstrate our explanation by the model calculations
utilizing Green's function integration. We have
shown~\cite{Kaza_JETPL_2006}, that under the following conditions
observed in our experiment: (a) the SPP wave on semi-infinite
surface is well described by the decayed sinewave (b) the sinusoidal
field distribution, produced by the straight edge of the metal mask,
starts as sharp as within 150-200 nm from the edge (c) there is no
significant reflections of SPP wave from the mask edges (the
conditions in fact take place), the quantitative description of the
SPP field $E^{\prime}_{spp}$ can be dramatically simplified. Namely,
$E^{\prime}_{spp}$ can be obtained in complex values by
2-dimensional integration of Green's function (see
e.g.~\cite{Zayats_GF_PR2005} )

\begin{equation}
E_{spp}(x,y)= \zeta \int \limits_{XY}
E_{las}(x^{\prime},y^{\prime})G(x-x^{\prime},y-y^{\prime})dx
^{\prime}dy^{\prime}
 \label{eq:8}
\end{equation}

where $E_{las}(x,y)=E_0exp(i\omega t+i\bf{kr})$ represents the
external light over the surface point $(x,y, z=0)$ with its
direction described by $\varphi$ and $\theta$, complex value $\zeta$
expressed from the field components continuity conditions represent
the field component above the point $(x,y)$ for the SPP wave created
at the point $(x',y')$. Once launched at the point $(x',y')$, the
SPP wave representing interaction of the point $(x',y')$ with all
other surface points, automatically fulfils the wave equations if
Green's function $G(x-x' ,y-y')$ is an eigenfunction of the
equations mentioned. The less is the "skin-layer depth" (light
penetration depth defined by (\ref{eq:sigma_z})), the better is
agreement between experimentally observed data and simulation
results in such a simplified model. For the description of exited
polarization delivery from one point of the surface to another,
cylindric functions~\cite{Hankel_Code_Book}, expressed in central
coordinates, are more convenient than plane waves. Nevertheless,
both kinds of functions are proven to be solutions of the wave
equation under the same conditions and assumptions.

\begin{figure}
\includegraphics[width=\columnwidth]{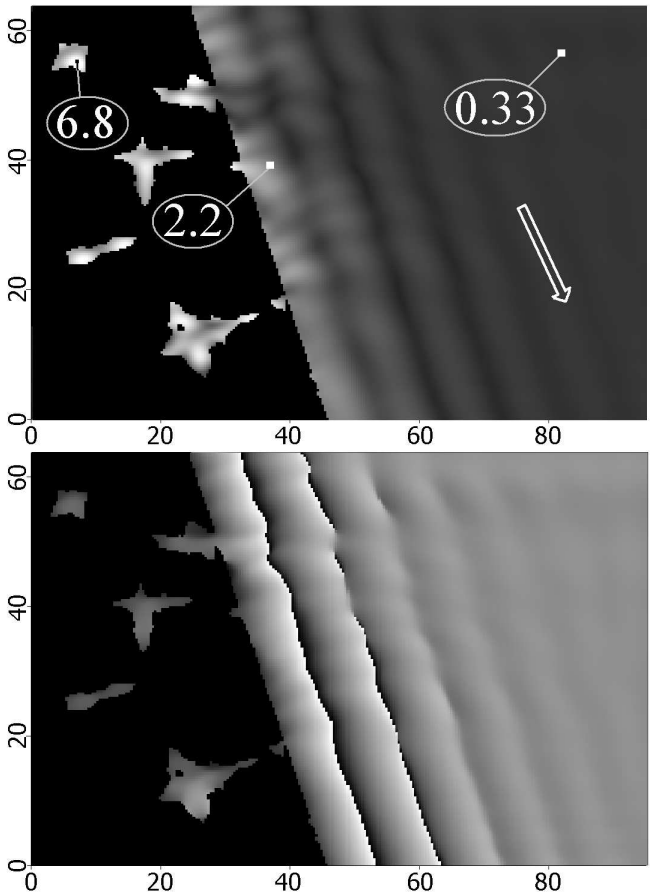}
\caption{\label{fig:theorMap}Excited E-field amplitude/phase map
calculated with Greens function integration. Light frequency is
$936~cm^{-1}$. Image size $95\times65~\mu m$. Light incidence angles
$\varphi=65^{\circ}$ (shown by arrow) and $\theta
^{\prime}=60^{\circ}$ from normal. }
\end{figure}

The results of numerical calculations are shown in
Fig.~\ref{fig:theorMap}. The shapes of the areas to be used for
integration were taken from bitmaps of real surface images. The SPP
wave pattern in every mask opening was calculated separately.
Despite the simplicity of the model used, the calculations still
show the following: (a) the amplitude over the mask edge and over
the SiC at any small hole in the mask is higher than over the
infinitely open surface, (b) the amplitude variations near the edge
are stronger if the illumination light comes from the metal mask
side and weaker for the opposite orientation of the sample, and also
(c) even the configuration of the standing wave in a small hole has
qualitative agreement with the observed scattering amplitude map for
different directions of the light.

Comparing present results with the results of previous
investigations of the similar samples, we can state the following.
The modification of the tip polarizability due to the interaction
with its dipole image in the surface is defined mainly by the
surface region corresponding to the tip radius and tip-sample
distance. The dramatic increase of the tip scattered near-field
signal was mentioned~\cite{Hilenbr_materContr_Nat_02} by comparison
of the scattered field amplitude while scanning over SiC surface at
the frequency close to SiC lattice resonance and the amplitude over
non-resonant surface of the metal mask. The mapping of the sample
dielectric function (which defines variations in
$\alpha^{(2\Omega)}_{eff}( \varepsilon_s(x,y))$, the tip effective
polarizability) was also
performed~\cite{Ocelic_CheckrBrd_Nat_Mat_04} to observe a material
contrast between two kinds of dielectric. In the current research,
instead, the dielectric function was assumed to be constant for
whole SiC surface being mapped, so that the sSNOM tip was considered
as just a tool (of constant $\alpha^{(2\Omega)}_{eff}(
\varepsilon_s)$ sensitivity factor) to scatter the surface field.
Thus, the scattering amplitude was compared not for the metal and
dielectric surface, but rather for the different areas of the same
dielectric (SiC) media.


Thus, we can conclude that, in the presence of efficient lateral
transfer of lattice polarization by SPP waves, the electromagnetic
response of resonant dielectric surface can not be described just by
the tailoring of local fields beneath and above the surface.
Instead, it must be considered as collective effect, with
integration over all surface points around. In the terms of
Green's-function formalism, explanation looks as: only those
neighbors contribute to the local surface polarization by
constructive way, which have no retardation. On the large open
surface the destructive contribution of the relatively far neighbors
is stronger because their amount is more, being proportional to the
distance $R$ from them $dS=2{\pi}RdR$. Thus, excluding such far
neighbors by covering the surface with a metal mask containing just
small openings, leads to significant enhancement of the local
amplitude.

Author thanks Dr.~R.Hillenbrand and Dr.~F.Keilmann for their
everyday's encouragement in the works reported and their very
constructive criticism. The author is indebted to use DSP
demodulation circuit programmed by N.Ocelic. The mechanical parts of
the sSNOM head work always perfectly due to the professional skill
of R.Gatz who fabricated it.


\end{document}